\documentclass[journal,onecolumn]{IEEEtran}
\usepackage{graphics}
\usepackage{amssymb}
\usepackage{graphicx}
\usepackage{fancybox}
\usepackage{amsmath}
\usepackage{subfigure}
\usepackage{algorithm}
\usepackage{algorithmic}
\usepackage{setspace}
\usepackage{booktabs}
\usepackage{caption}
\usepackage{wrapfig}
\usepackage{url}
\usepackage{dirtytalk}
\usepackage[normalem]{ulem}
\useunder{\uline}{\ul}{}

\begin{document}
\twocolumn[{%
\vspace{20mm}
\renewcommand\twocolumn[1][]{#1}%
{ \large
\begin{itemize}[]

\item[\textbf{Citation}]{G. AlRegib, M. Prabhushankar, K. Kokilepersaud, P. Chowdhury, Z. Fowler, S. Trejo Corona, L. Thomaz, A. Majumdar,  ``Ophthalmic Biomarker Detection: Highlights from the IEEE Video and Image Processing Cup 2023 Student Competition," in \textit{IEEE Signal Processing Magazine,} 2024.}

\item[\textbf{Review}]{Date of Acceptance: May 12 2024}

\item[\textbf{Codes}]{\url{https://github.com/olivesgatech/VIPCUP2023_OLIVES.git}}

\item[\textbf{Bib}]  {@ARTICLE\{alregib2024ophthalmic,\\
    author=\{G. AlRegib, M. Prabhushankar, K. Kokilepersaud, P. Chowdhury, Z. Fowler, S. Trejo Corona, L. Thomaz, A. Majumdar\},\\
    journal=\{IEEE Signal Processing Magazine\},\\
    title=\{Ophthalmic Biomarker Detection: Highlights from the IEEE Video and Image Processing Cup 2023 Student Competition\},\\
    year=\{2024\}\}}
\item[\textbf{VIP}]{\url{https://alregib.ece.gatech.edu/2023-vip-cup/}}
\item[\textbf{Contact}]{
olives.gatech@gmail.com\\\url{https://ghassanalregib.info/}\\}

\end{itemize} }}]
\newpage

\title{Ophthalmic Biomarker Detection: Highlights from the IEEE Video and Image Processing Cup 2023 Student Competition}
\author{Ghassan AlRegib, Mohit Prabhushankar, Kiran Kokilepersaud, Prithwijit Chowdhury, Zoe Fowler, Stephanie Trejo Corona, Lucas Thomaz, and Angshul Majumdar}

\maketitle
\begin{spacing}{1.5}
\section{Introduction}
Ophthalmic clinical trials study treatment efficacy and are performed with a specific purpose and set of procedures. %predetermined before trial initiation. 
Generally, this purpose is to assess one of two variables. The first variable is the control biomarker or other measurement that provides an assessment of the presence, absence, or severity of the studied disease. Treatment administration is then determined based on this control. The second variable is the treatment protocol or regimen, which can be influenced by multiple factors such as drug dosage and type of drug used. 
%The time between treatment, the drug used for treatment and its dosage are among several factors that influence this protocol. %In general, it is desirable to have treatments that are less frequent to decrease any side effects patients may experience. However, frequency must not affect treatment efficacy. 
In fact, the variety and amount of data collected throughout the clinical trial process can also guide the treatment regimen. For example, 1D clinical measurements and structural biomarkers are commonly observed, as well as 2D color and near-IR fundus and even 3D optical coherence tomography (OCT). Thus, clinical trials are designed to identify which of the aforementioned data modalities are relevant. However, because of the presence of a large quantity of these multifarious data, it is impractical to conduct trials by controlling each of these variables.

\begin{figure*}[h]
\centerline{\includegraphics[width=5in]{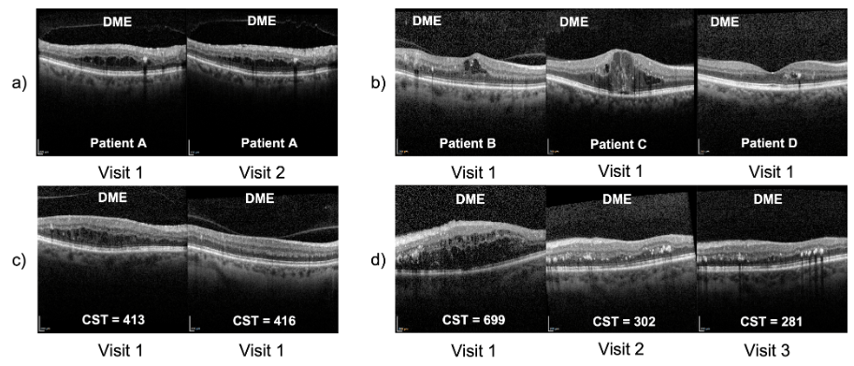}}
\caption{An illustration of personalization challenges within the dataset.}
\label{OCT}
\end{figure*}

Deep learning-based artificial intelligence systems have shown an affinity for understanding and exploiting multi-modal correlations in a high dimensional space. The {\fontfamily{qcr}\selectfont OLIVES} dataset \cite{prabhushankar2022olives} is a multi-modal dataset for deep learning applications, providing access to 2D and 3D ophthalmic imaging data, scalar clinical labels, and biomarkers for two different eye diseases. Prior results utilizing this dataset in \cite{prabhushankar2022olives} use state-of-the-art self-supervised methods on assessing the presence of biomarkers, though results varied between biomarkers. This variance is due to the presence and absence of biomarkers being a personalization challenge rather than a generalization challenge. This phenomenon is demonstrated in Figure \ref{OCT}, where the variation within OCT scans of patients between visits can be minimal while the difference in manifestation of the same disease across different patients may be substantial. The domain difference between OCT scans can arise due to pathology manifestation across patients (Fig.1a and Fig.1b), clinical labels (Fig.1c), and the visit along the treatment process when the scan is taken (Fig.1d). Thus, while deep learning algorithms have shown an affinity towards generalization, they are lacking in personalization.

Due to prior deep learning approaches' inability to balance generalization and personalization, we use the {\fontfamily{qcr}\selectfont OLIVES} dataset to host the 2023 Video and Image Processing (VIP) Cup to further explore and address this issue in biomarker detection of ophthalmology data. %this challenge aims to address the issue of personalization.  for the task of biomarker detection in ophthalmology data
The VIP Cup offers a unique experience to undergraduates, allowing students to work together to solve challenging, real-world problems with video and image processing techniques. Each team includes one faculty member serving as an advisor, at most one graduate student as a mentor, and at least three but no more than ten undergraduate students. Formed teams participate in an open competition, with the top three teams presenting their work at the final competition held at the 2023 IEEE International Conference on Image Processing (ICIP) in Kuala Lumpur, Malaysia. Travel costs of finalist teams are supported by the IEEE Signal Processing Society (SPS). In this article, we share an overview of the VIP Cup experience including competition tasks, participating teams, technical approaches, statistics, and competition experience.

\section{Tasks and Evaluation Criteria}
The 2023 VIP Cup competition consists of two distinct phases, each with unique challenges and assessment criteria. In this section, we provide an overview of the tasks teams are challenged to complete, along with the provided resources and the criteria teams are evaluated on.

\subsection{VIP Cup Tasks}

The competition takes place across two phases, which aim at testing teams' approaches to both the generalization and personalization challenges present within ophthalmology data. Throughout this section, we describe the specific tasks present within each competition phase.

\begin{table*}[]
\centering
\begin{tabular}{|l|l|l|l|l|l|}
\hline
\textbf{Modality} & \textbf{Per Visit} & \textbf{Per Eye} & \textbf{Train Total} & \textbf{Phase 1 Test Set} & \textbf{Phase 2 Test Set} \\ \hline
OCT               & 49                 & $N_{p}*49$            & 78189                & 3871                      & 250                       \\ \hline
Clinical          & 4                  & $N_{p}*4$             & 5082                 & 320                       & 250                       \\ \hline
Biomarker         & 6                  & $N_{p}*49*6$          & 469134               & 23226                     & 1500                      \\ \hline
\end{tabular}
\caption{Provided datasets statistics across three modalities (OCT, clinical, and biomarker data), where $N_{p}$ indicates the total number of patients}
\label{tab:olives}
\end{table*}
% \begin{table*}[]
% \centering
% \begin{tabular}{|l|l|l|l|l|l|}
% \hline
% \textbf{Modality} & \textbf{Per Visit} & \textbf{Per Eye} & \textbf{Train Total} & \textbf{Phase 1 Test Set} & \textbf{Phase 2 Test Set} \\ \hline
% OCT               & 49                 & $N_{p}*49$            & 78189                & 3871                      & 250                       \\ \hline
% Clinical          & 4                  & $N_{p}*4$             & 5082                 & 320                       & 250                       \\ \hline
% Biomarker         & 6                  & $N_{p}*49*6$          & 469134               & 23226                     & 1500                      \\ \hline
% \end{tabular}
% \caption{Provided datasets statistics across three modalities (OCT, clinical, and biomarker data), where $N_{p}$ indicates the total number of patients}
% \label{tab:olives}
% \end{table*}

\subsubsection{Open Competition Phase 1 - Generalization Task}

The overall task in Phase 1 was to predict the presence or absence of six different biomarkers on every OCT scan in the provided test set. This task, in general, can be challenging due to different biomarkers exhibiting varying levels of granularity, where some biomarkers are more visually apparent than others. 

The training set for this task was derived from the {\fontfamily{qcr}\selectfont OLIVES} dataset and consisted of the labeled OCT scans, along with the associated clinical information and the corresponding biomarker labels as the ground truth. Participants were encouraged to make use of all available modalities.  Additionally, the participants were provided with a testing dataset that was derived from a recent clinical study in collaboration with Retina Consultants of Texas (Houston, TX, USA). However, teams were only provided with these OCT scans and their clinical information. 
%Thus, participants used all available data to predict the biomarkers associated with each OCT scan in this test set. 
Once teams had developed their approach, they submitted their algorithms to Codalab for evaluation.

\subsubsection{Open Competition Phase 2 - Personalization Task}

In Phase 1 of the competition, each slice in the test set was treated as its own independent entity. In reality, however, every set of 49 slices within the test set is associated with a specific patient’s eye. Hence, practitioners may be interested in performance with respect to the patient’s eye as a whole, rather than performance with respect to isolated slices of the retina. Thus, in Phase 2 of the competition the organizers better assess how well the model is able to personalize. 

For Phase 2, the organizers introduce a new held-out test set that replaces the previous test set from the first stage of the competition. This new test set is labeled by our medical partners from Retina Consultants of Texas in the same manner as the first stage, with every image associated with 6 biomarkers. In the Phase 1 test set, a large pool of images are drawn from different visits of the same patient. Hence, Phase 1 test set has substantial redundancy even with a larger number of images, as evidenced by Table~\ref{tab:olives}. 
%Furthermore, the Phase 1 test set was drawn from a clinical trial with similar population demographics to that of the training set. Both of these considerations limit the ability of the original test set to rigorously test the generalizability of the models created in this competition. 
The Phase 2 test set, on the other hand, is drawn from a larger patient pool and population base which allows a better assessment of both the generalizability and personalization aspects of the submitted algorithms. To perform the personalization aspect of the competition, the registered teams from Phase 1 were given the opportunity to re-train their models and submit the biomarker prediction files for each image in the test set. The overall finalists were then drawn from these performance outcomes on the Phase 2 test set.

\subsection{Resources}

%\subsubsection{Datasets and Code}

Teams were provided with access to the labeled {\fontfamily{qcr}\selectfont OLIVES} dataset and another ophthalmology clinical trial dataset from Retina Consultants of Texas for Phase 1, as well as initial Python code using this dataset \cite{prabhushankar2022olives}. In Phase 2, we provide a subset of Kermany dataset for training purposes \cite{kermany2018identifying}. We list a summary of the overall amount of multimodal data included for Phase 1 and Phase 2 in Table \ref{tab:olives}, detailing information about the amount of data included in both train and test sets. The difference in challenge tasks is clearly indicated in this table. For example, in Phase 2, our new held-out test set contained biomarker labels from the Kermany dataset labeled privately by our medical partners \cite{kermany2018identifying}. This difference between Phase 1 and Phase 2 test sets is demonstrated in Table \ref{tab:olives}, where the Phase 2 test set has a much larger patient diversity (167 vs. 40 unique patients) while having fewer images (250 vs. 3871). In addition, teams were provided with potential suggestions for improving detection, such as using integrating clinical values into the designed algorithm as recently done in contrastive learning literature using OCT data \cite{kokilepersaud2022clinical}, \cite{kokilepersaud2022gradient}.

\subsection{Evaluation Criteria}
We implement two distinct evaluation approaches to judge the teams' generalization and personalization approaches in Phase 1 and 2. The top three teams at the end of Phase 2 are then evaluated to determine the final rankings. In this section, we go over the evaluation criteria for each phase, as well as the criteria the finalists are judged on.

\subsubsection{Phase 1 Evaluation}

For Phase 1, we draw the test set from an entirely separate clinical trial with the same disease pathology as the training set, where each slice in the test set is treated as its own independent entity in the Phase 1 biomarker detection task. Thus, we make use of the macro averaged F1-score to measure the performance for this phase of the competition. This metric is the equally weighted average of the F1-scores for each of the biomarkers in the 16 biomarker classification task. This metric is desirable because the test set is highly imbalanced, making it impossible to guarantee an equal distribution of biomarkers within each OCT scan. The F1 score is more sensitive to this dataset imbalance, as it is the harmonic mean of precision and recall. In addition, we want our metric to treat each class of biomarker as equally important, rather than reflecting a bias towards classes with more instances present. Therefore, we will designate the top teams as those with the highest macro-averaged F1-score across all 6 biomarkers. With the starter code provided, we achieve a baseline score of 0.6315 when using the described metric for the Phase 1 dataset.

\subsubsection{Phase 2 Evaluation}

For Phase 2, the test set is from a general population setting and not from a clinical trial. 
%The number of patients are $167$, four times the number in Phase 1 test set. 
For this portion of the competition, we compute the macro F1-score for each slice in the same way as Phase 1. Unlike Phase 1, however, we will now average the F1-scores with respect to each set of slices associated with an individual patient, rather than averaging across the test set as a whole. This will result in an associated macro F1-score with respect to each individual patient.
%Teams are ranked with respect to performance on each of the patients in the test set. 
The overall ranking of each team is established by averaging the scores across all patients. The three teams with highest performance in Phase 2 are then invited to present their work in the final phase at ICIP 2023. With the provided starter code, we achieve a baseline result of 0.6943 using the described evaluation metric for the Phase 2 data.

\subsubsection{Finalist Judging Criteria}

Judging for the final phase of the competition was held live at the ICIP 2023 conference. It was based on five equally weighted criteria. Each of the three finalist teams are scored on the five criteria and the team with the highest score places 1st, the team with the second highest score places 2nd, and the team with the third highest score places 3rd in the competition. The five equally weighted criteria are: (1) Innovation of the proposed approach, (2) Performance of the team on generalization metric, (3) Performance of the team on personalization metric, (4) Quality and clarity of the final report, and (5) Quality and clarity of the presentation. Each criterion is scored with a 1, 2, or 3; the best team in each criterion receives 3 points, the second-best team receives 2 points, and the third best team receives 1 point. The final winning rankings are based on the highest points awarded from the five criteria during judge deliberations at the end of the competition. Final rankings are ultimately decided by the judges at their discretion.

\section{Participation Statistics}
% - Need: graphs maybe to highlight countries that participated by percentage\\
% - how many teams total
% - Might be interesting: did teams/universities that compete this year also compete in previous years?
% \begin{wrapfigure}{r}{0.3\textwidth}
% \begin{center}
%     \includegraphics[width=0.45\textwidth]{Images/fig2.png}
% \end{center}
% \caption{Daily submission count over the course of the competition}
% \label{participation}
% \end{wrapfigure}

\begin{figure}[h!]
    \centering
    \includegraphics[width=0.8\linewidth]{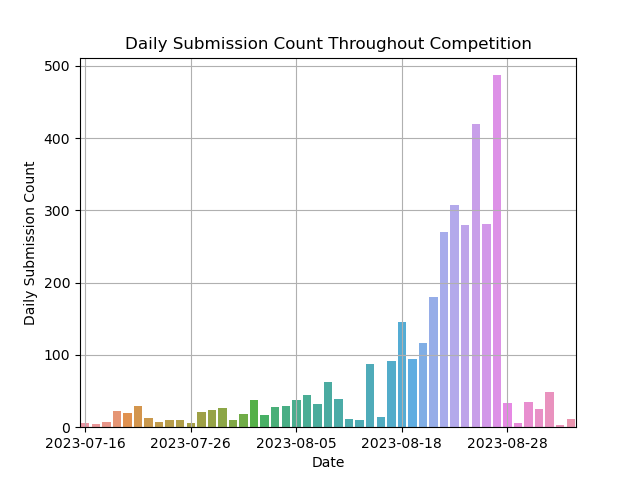}
    \caption{Daily submission count over the course of the competition}
    \label{participation}
\end{figure}

The 2023 VIP CUP experienced an unprecedented surge in participation, demonstrating its widespread appeal and global engagement. Hosted on Codalab, the competition garnered an astounding total of 3,501 submissions, reflecting the remarkable enthusiasm and dedication of the participants. The competition witnessed an average of 80 daily submissions, underscoring the sustained interest and commitment throughout its duration. In total, the event attracted 17 registered teams using ICIP's Conference Management Services (CMS). The vibrant participation was further highlighted by the involvement of 121 individual participants, showcasing the diverse and competitive nature of the field. We highlight daily submission statistics gathered from Codalab in Figure \ref{participation} throughout the competition. Notably, the global reach of the VIP Cup 2023 was evident, with significant representations from countries such as Bangladesh, India, Sri Lanka, and China, emphasizing the competition's truly international character.

%The overwhelming response underscores the success of the event in fostering collaboration and skill development within the global community.
% \begin{figure}[h]
% \centerline{\includegraphics[scale=.5]{Images/VIP_Participation.png}}
% \caption{Daily submission count over the course of the competition}
% \label{participation}
% \end{figure}

% stats from codalab
% 2 different websites; combine info
\section{Finalists}
The finalist teams of the 2023 VIP Cup, as well as their respective rankings are as follows:

\textbf{Team Synapse} \textit{(first place)}
\begin{itemize}
    \item \textit{Affiliation:} Bangladesh University of Engineering and Technology
    \item \textit{Students:} H.A.Z. Sameen Shahgir, Khondker Salman Sayeed, Tanjeem Azwad Zaman, Md. Asif Haider
    \item \textit{Mentor:} Sheikh Saifur Rahman Jony
    \item \textit{Supervisor:} M. Sohel Rahman
\end{itemize}
% \begin{wrapfigure}{r}{0.3\textwidth}
% \begin{center}
%     \includegraphics[width=0.3\textwidth]{Images/fig3.jpg}
% \end{center}
% \caption{Team IITH with the judging panel at the VIP Cup 2023 competition.}
% \label{iith-comp}
% \end{wrapfigure}

\textbf{Team Neurons} \textit{(second place)}
\begin{itemize}
    \item \textit{Affiliation:} Bangladesh University of Engineering and Technology
    \item \textit{Students:} Md. Touhidul Islam, Md. Abtahi Majeed Chowdhury, Mahmudul Hasan, Asif Quadir
    \item \textit{Supervisor:} Lutfa Akter
\end{itemize}

\textbf{Team IITH} \textit{(third place)}
\begin{itemize}
    \item \textit{Affiliation:} Indian Institute of Technology, Hyderabad
    \item \textit{Students:} Aaseesh Rallapalli, Lokesh Badisa, Nithish S, Utkarsh Doshi
    \item \textit{Supervisor:} Soumya Jana\\
\end{itemize}

\begin{figure}[h!]
\centerline{\includegraphics[scale=.04]{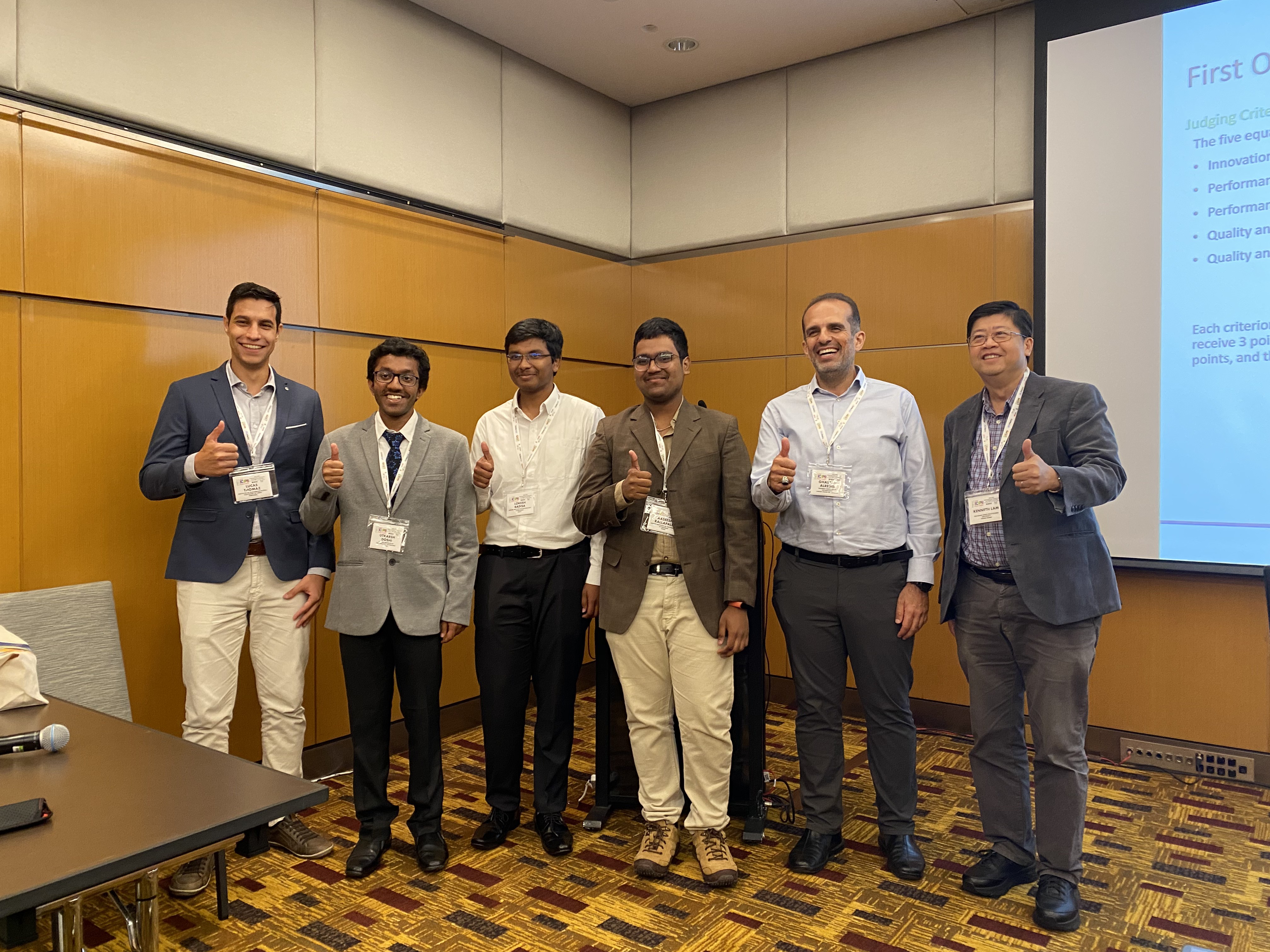}}
\caption*{Team IITH with the judging panel at the VIP Cup 2023 competition.}
\label{iith-comp}
\end{figure}
\section{Highlights of Technical Approaches}
%- Need: go through finalists' approaches and document that\\
%- Add: Generalization / personalization graphs from presentation, patient performance ranking, generalization ranking, all F1 scores tables

The finalists for the 2023 VIP CUP Competition are teams Neurons, IITH, and Synapse. In the development of each approach, all teams noted that they had to come to terms with resolving the discrepancy between local and global structures in the provided data. This problem for the associated OCT imaging data is discussed extensively in \cite{kokilepersaud2023clinically} where the authors discuss how algorithms must account for biomarkers with both fine and coarse granularity. The challenge in medical data is that both types of biomarkers can appear simultaneously and models are more likely to identify the coarse biomarkers while showing poor performance on the fine-grained biomarkers. 
%We complicate this issue further by having participants consider this problem within the context of both overall generalization as well as within a patient-specific personalization metric. 
Below, we summarize each approach and specifically highlight, where appropriate, attempts by each team of addressing this granularity issue. Afterwards, we analyze how their differences manifest across performance metrics related to both the personalization and generalization tasks of this competition. 

\paragraph{Team IITH} surprisingly showed that just an Inception v3 \cite{szegedy2015going} backbone with sufficient hyperparameter tuning was enough to obtain competitive performance. The participants arrived at this conclusion after a wide architecture search across a variety of convolutional and transformer based architectures. The team notes that specialized architectural considerations in Inception v3 such as factorizing convolutions, spatially separable convolutions, and efficient grid reduction may enable this architecture to perform feature extraction at multiple scales that is optimal for the biomarker detection task.  

\begin{table*}[]
\centering
\begin{tabular}{cccccccr}
\hline
\multicolumn{8}{|c|}{\textbf{Phase 1 Biomarker-Specific Generalization F1-Score}}                                                                                                                                                                                                             \\ \hline
\multicolumn{1}{|c|}{\textbf{Team}} & \multicolumn{1}{c|}{\textbf{B1}} & \multicolumn{1}{c|}{\textbf{B2}} & \multicolumn{1}{c|}{\textbf{B3}} & \multicolumn{1}{c|}{\textbf{B4}} & \multicolumn{1}{c|}{\textbf{B5}} & \multicolumn{1}{c|}{\textbf{B6}} & \multicolumn{1}{c|}{\textbf{Overall}} \\ \hline
Elemenopi                           & 0.656                            & 0.756                            & 0.784                            & 0.832                            & 0.852                            & \textbf{0.992}                   & 0.7518                                \\
IITH                                & 0.684                            & 0.74                             & 0.732                            & 0.94                             & 0.864                            & \textbf{0.992}                   & 0.747                                 \\
MEA                                 & 0.38                             & 0.62                             & 0.592                            & 0.636                            & 0.156                            & 0.02                             & 0.197                                 \\
Neurons                             & 0.648                            & 0.76                             & 0.788                            & 0.96                             & 0.768                            & 0.984                            & 0.779                                 \\
Optimpus                            & 0.688                            & 0.488                            & 0.512                            & 0.776                            & 0.852                            & \textbf{0.992}                   & 0.646                                 \\
Pixel Pulse                         & 0.652                            & 0.664                            & 0.656                            & 0.912                            & 0.788                            & 0.564                            & 0.532                                 \\
Pixel Vision                        & 0.476                            & 0.648                            & 0.6                              & 0.648                            & 0.16                             & 0.072                            & 0.398                                 \\
Sharks                              & 0.652                            & 0.696                            & 0.632                            & 0.616                            & 0.848                            & \textbf{0.992}                   & 0.668                                 \\
Source Code                         & 0.676                            & 0.552                            & 0.492                            & 0.848                            & \textbf{0.884}                   & 0.916                            & 0.6133                                \\
Spectrum                            & 0.716                            & 0.74                             & 0.748                            & 0.908                            & 0.804                            & 0.968                            & 0.7504                                \\
Synapse                             & 0.712                            & \textbf{0.776}                   & \textbf{0.836}                   & \textbf{0.956}                   & 0.86                             & \textbf{0.992}                   & \textbf{0.806}                        \\
Tesseract                           & 0.732                            & 0.676                            & 0.592                            & 0.932                            & 0.848                            & \textbf{0.992}                   & 0.644                                 \\
Ultrabot AIO                        & 0.732                            & 0.632                            & 0.604                            & 0.92                             & 0.856                            & 0.976                            & 0.721                                 \\
UNNC\_ISEAN                         & \textbf{0.76}                    & 0.74                             & 0.776                            & 0.136                            & 0.82                             & 0.132                            & 0.568                                 \\
UNNC\_POWER                         & 0.72                             & 0.728                            & 0.756                            & 0.536                            & \textbf{0.884}                   & 0.036                            & 0.543                                 \\ \hline
\multicolumn{8}{|c|}{\textbf{Phase 2 Biomarker-Specific Personalization F1-Score}}                                                                                                                                                                                                            \\ \hline
\multicolumn{1}{|c|}{\textbf{Team}} & \multicolumn{1}{c|}{\textbf{B1}} & \multicolumn{1}{c|}{\textbf{B2}} & \multicolumn{1}{c|}{\textbf{B3}} & \multicolumn{1}{c|}{\textbf{B4}} & \multicolumn{1}{c|}{\textbf{B5}} & \multicolumn{1}{c|}{\textbf{B6}} & \multicolumn{1}{c|}{\textbf{Overall}} \\ \hline
\textbf{IITH}                       & 0.6893                           & 0.7528                           & 0.7578                           & 0.9483                           & \textbf{0.8516}                  & \textbf{0.988}                   & \textbf{0.8215}                       \\
\textbf{Neurons}                    & 0.6397                           & 0.7762                           & 0.7891                           & \textbf{0.9657}                  & 0.7684                           & 0.9847                           & \textbf{0.8116}                       \\
\textbf{Synapse}                    & 0.7173                           & \textbf{0.7946}                  & \textbf{0.8525}                  & 0.9632                           & 0.8456                           & \textbf{0.988}                   & \textbf{0.8527}                       \\
\hline
Elemenopi                           & 0.6527                           & 0.7734                           & 0.8122                           & 0.817                            & 0.8396                           & \textbf{0.988}                   & 0.7966                                \\
MEA                                 & 0.3861                           & 0.6236                           & 0.592                            & 0.636                            & 0.172                            & 0.0299                           & 0.3822                                \\
Optimpus                            & 0.6897                           & 0.5087                           & 0.5207                           & 0.7902                           & 0.8506                           & 0.988                            & 0.697                                 \\
Pixel Pulse                         & 0.6477                           & 0.6709                           & 0.6954                           & 0.9183                           & 0.7926                           & 0.556                            & 0.6984                                \\
Pixel Vision                        & 0.4774                           & 0.6466                           & 0.6188                           & 0.6903                           & 0.1727                           & 0.0689                           & 0.4139                                \\
Sharks                              & 0.647                            & 0.6889                           & 0.6282                           & 0.6191                           & 0.8306                           & \textbf{0.988}                   & 0.7056                                \\
Source Code                         & 0.7028                           & 0.587                            & 0.4977                           & 0.8627                           & 0.8849                           & 0.9254                           & 0.7274                                \\
Spectrum                            & 0.7326                           & 0.7518                           & 0.7505                           & 0.9068                           & 0.8038                           & 0.9676                           & 0.8067                                \\
Tesseract                           & 0.744                            & 0.6928                           & 0.6158                           & 0.9378                           & 0.8336                           & \textbf{0.988}                   & 0.7921                                \\
Ultrabot AIO                        & 0.7421                           & 0.6364                           & 0.5963                           & 0.9311                           & 0.8425                           & 0.9757                           & 0.7723                                \\
UNNC\_ISEAN                         & \textbf{0.771}                   & 0.7658                           & 0.7997                           & 0.1379                           & 0.8086                           & 0.1349                           & 0.5426                                \\
UNNC\_POWER                         & 0.7178                           & 0.7464                           & 0.7897                           & 0.5241                           & 0.8739                           & 0.0479                           & 0.593                                
\end{tabular}
\caption{The F1-Scores for all 15 teams on individual biomarkers across both phases of the competition are shown above. For Phase 2, the final three teams (IITH, Neurons, and Synapse) are bolded.}
\label{tab:biomarker}
\end{table*}

\paragraph{Team Neurons} used their signal and image processing knowledge to understand that feature extraction is an important part of classification, particularly due to how visually different the biomarkers appear \cite{islamophthalmic}. Hence, they made the insight that convolutional networks (CNN) are good at local feature extraction with their fixed kernels, while transformers use attention to extract features from a global context. Therefore, using an ensemble of both enables extraction of both types of features. Specifically, the used an EfficientNet \cite{tan2019efficientnet} architecture for local feature extraction and a Max-ViT \cite{tu2022maxvit} transformer for global understanding. They then split the overall OLIVES dataset based on TREX or PRIME clinical trials specifically. The ensemble was then formed by training their models on the subsets TREX + PRIME, TREX only, and PRIME only. This allows different subsets of the models to be tuned to the different types of features that each trial provides. During inference, each pair of models trained on each subset to produce a prediction that are then used in an ensemble fashion to produce the final decision.

\paragraph{Team Synapse} similarly used their image and signal processing knowledge to inform their observation that an ensemble of models with one performing local feature extraction and one performing global attention is optimal for this task due to biomarkers of varying granularity \cite{shahgirophthalmic}. However, in contrast to team Neurons, they used a vision transformer with a convolutional block as the local feature extractor, rather than a CNN alone. In this case, Max-VIT \cite{tu2022maxvit} was the local extractor and EVA-02 \cite{fang2023eva} performed global attention. The team notes that the the use of normal attention compared to strided attention causes the global vs. local feature extraction capability. An ensemble was then created by having Max-VIT predict only fine-grained biomarkers, while EVA-02 was responsible for coarse-grained biomarkers.

A closer look at the performance of each approach in a variety of settings provides deeper insight into the personalization and generalization capabilities of each algorithm. In Table \ref{tab:biomarker}, we note the performance of all teams on individual biomarkers, where the highest performance per biomarker is denoted in bold. All teams performed similarly well on B4, B5, and B6, but a substantial gap exists on B1, B2, and B3. Team Synapse performed better on the more difficult fine-grained biomarkers compared to either of the other two finalists. This trend holds for both the personalization and generalization performance.
%and suggests that their approach was able to rectify local and global features to a greater extent. 

\begin{figure}[h]
\centerline{\includegraphics[scale=.53]{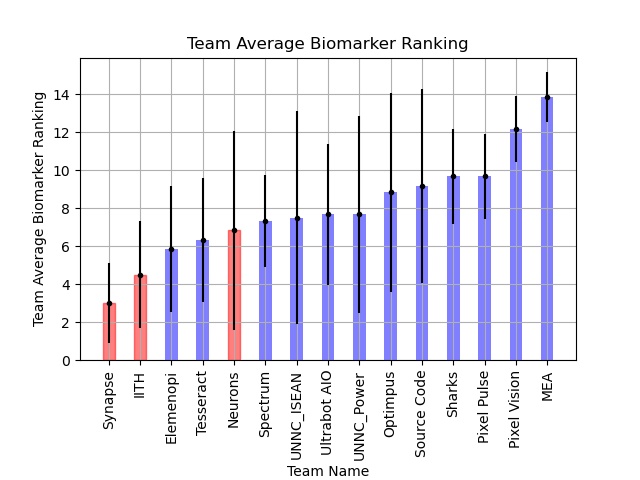}}
\caption{This figure shows the average and associated standard deviation of the ranking of each team on each biomarker with first being the best. The top three teams overall are highlighted in red.}
\label{biomarker_ranking}
\end{figure}

\begin{figure}[h]
\centerline{\includegraphics[scale=.53]{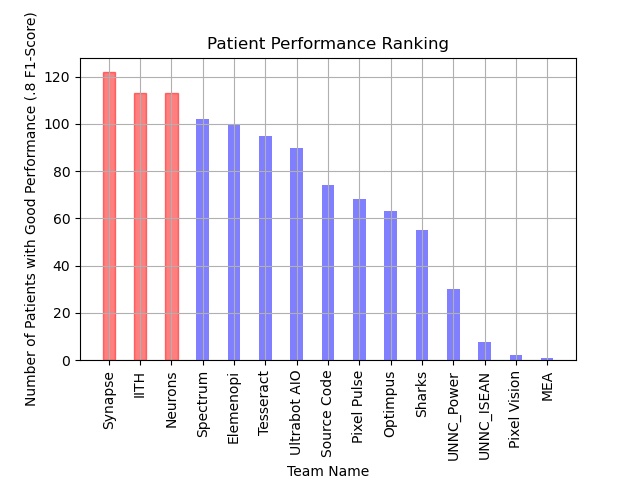}}
\caption{This shows the number of patients on the test set where the performance exceeded a good F1-Score threshold (.8) for each team in the competition. The top 3 teams are highlighted in red.}
\label{patients_ranking}
\end{figure}

\begin{figure}[h]
\centerline{\includegraphics[scale=.53]{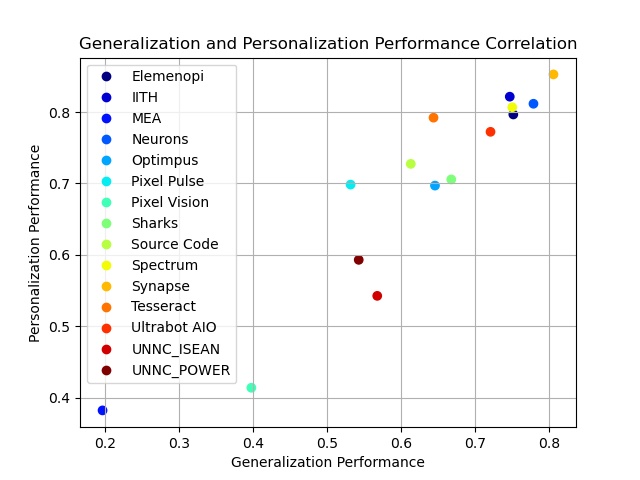}}
\caption{This plot shows the correlation shows every team's generalization and personalization performance.}
\label{generalization_personalization}
\end{figure}

\begin{figure}[h]
\centerline{\includegraphics[scale=.53]{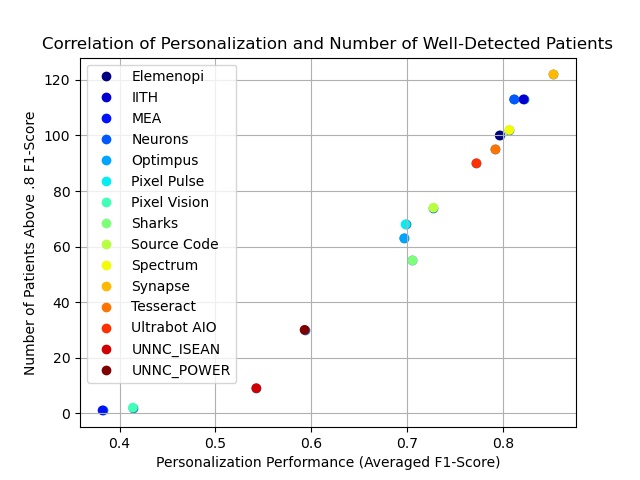}}
\caption{This plot shows how well every team's personalization performance correlates with the number of patients that have a good (.8 or higher) associated F1-Score.}
\label{personal_ranking}
\end{figure}

We also ranked each team in Figure \ref{biomarker_ranking} based on their generalization performance on each biomarker. Team Synapse and IITH performed consistently well across all biomarkers. Team Neurons exhibited higher variation as shown by better performance on specific biomarkers, but worse on others. We also ranked teams based on the number of patients where performance crossed a satisfactory F1-score threshold (0.8) in Figure \ref{patients_ranking}. 
%All the top-performing teams had the most number of patients where their performance was above this threshold as shown in Figure \ref{patients_ranking}. 
In Figures \ref{generalization_personalization} we show a scatter plot that demonstrates how correlated performance on the generalization and personalization tasks are for each team. 
%Overall, it seems that doing well on one task transfers to doing well on the other. 
We observe that personalization performance correlates with performing well on a higher number of patients as shown in Figure \ref{personal_ranking}. In both plots, the top 3 teams clustered in the high performing top-right region of the figures. Additionally, this comes with a noticeable separation from the overall winner Team Synapse. 
\section{Insights and Takeaways}
\begin{figure}[h]
\centerline{\includegraphics[width=0.45\textwidth]{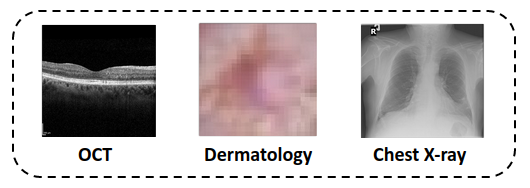}}
\caption{Difference in noise characteristics between OCT, dermatology, and chest x-ray images}
\label{data}
\end{figure}

In this section, we discuss insights used to create the VIP Cup 2023 challenge and the issues faced throughout the competition. From an organization standpoint, the topic of integrating artificial intelligence (AI) within medicine is one that has gained much attention within the research community within recent years and presents an excellent avenue towards introducing students to real-world applications of artificial intelligence. Ophthalmology, in particular, has been described as the first clinical application domain that has the potential for full automation \cite{beede2020human}. This is primarily because OCT and fundus photography are oftentimes easier to derive features from due to easier noise characteristics compared to other modalities of medical imaging tasks, making this type of data more amenable to AI algorithms that rely on deriving standard features to distinguish different classes. This is demonstrated in Figure \ref{data}, where the structures present in OCT are clearly visible, as opposed to the images taken from a dermatology (DermMNIST) and chest x-ray dataset (PadChest) \cite{yang2023medmnist}, \cite{bustos2020padchest}.

Another advantage of using OCT data is that it is associated with a clinical trial structure, rather than as isolated instances~\cite{fowler2023clinical}. The clinical trial structure enables the existence of many different modalities such as textual documents, clinical values, and other imaging modalities~\cite{logan2022decal}. Therefore, students get exposure to the existence and interaction of these different axes within the medical domain. Furthermore, the clinical trial structure acts as a regularizer in terms of the scope of the data they have access to. In other words, the size of the data they have access to is limited by the resources of the clinical trial thus preventing the existence of too large of a data source for students to work with effectively. This is important within the context of the VIP Cup such that the competition is accessible to students from different regions and backgrounds who may have lesser access to computational sources than others. Additionally, the clinical trial structure ensures that data is composed from wider demographic sources, which in turn ensures that the solutions devised by the students is indeed generalizable.

%Another major aspect is to get students used to the idea of handling data and performing some type of preprocessing pipeline before training models. OCT data is useful for this task because students have to consider whether some noise-reduction or edge enhancement strategy is useful before engaging in complete training.

\begin{figure}
\centerline{\includegraphics[scale=.058]{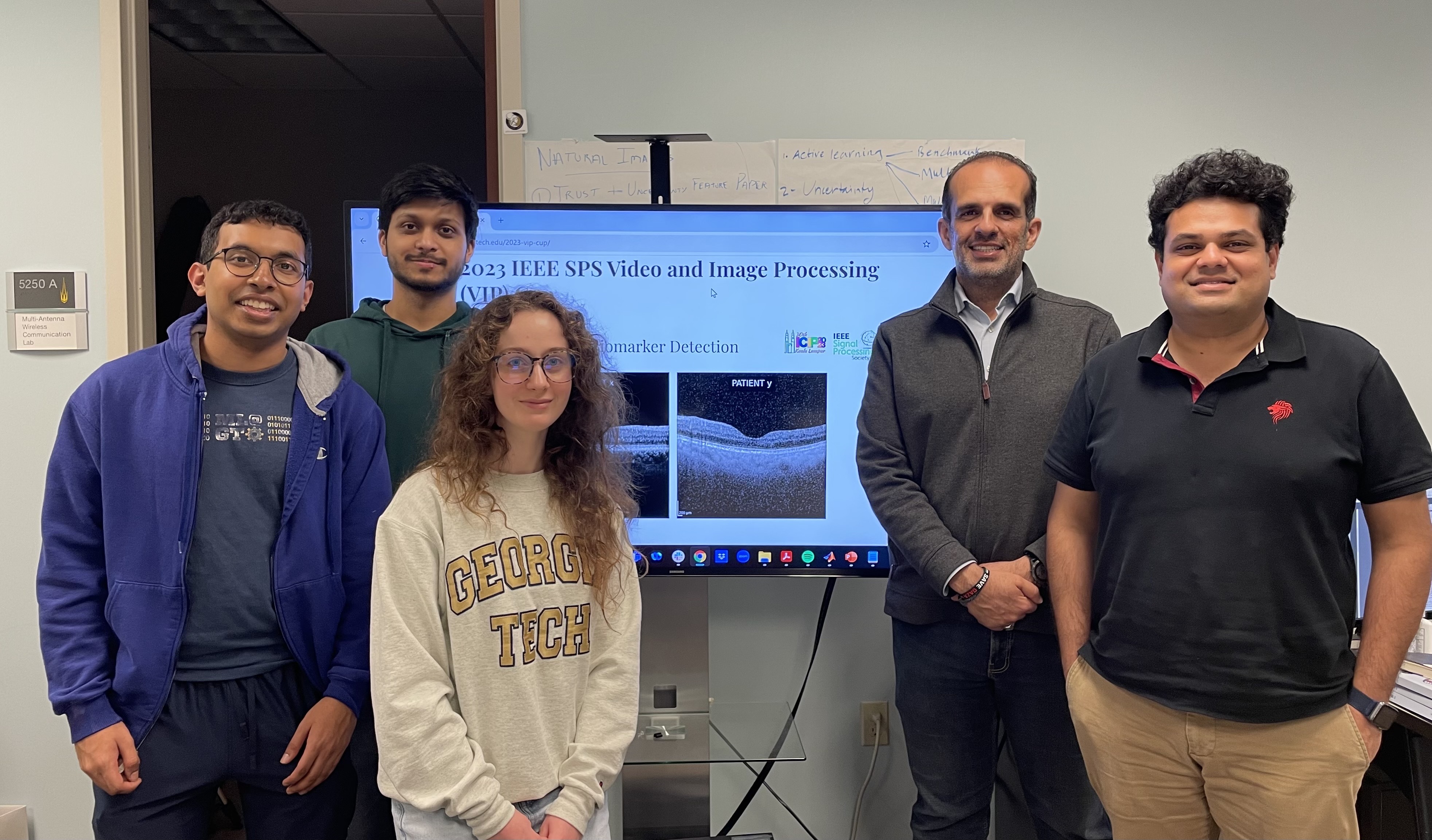}}
\caption{Organizing committee from the Olives lab at Georgia Institute of Technology. From left to right: Kiran Kokilepersaud, Prithwijit Chowdhury, Zoe Fowler, Ghassan AlRegib, and Mohit Prabhushankar.}
\label{olives-team}
\end{figure}

We now discuss the competition from an organization standpoint, as observed by the organizing committee (Figure \ref{olives-team}). When running the VIP Cup competition, we initially utilized Codalab for our project needs, which had automatic migration of Phase 1 submissions for hidden Phase 2 evaluations. However, we encountered performance and accessibility issues. Due to the high volume of traffic and migration of submissions on the Codalab platform, the server crashed. To combat this problem, we established a mirror site, where the submission resulted in a direct scoring without need for migration. This change drastically improved performance and reliability. Submissions were seamless with faster scoring, proving that the establishment of the mirror site was an efficient solution. This observation was corroborated by participating teams, as one team noted that for future VIP Cup iterations a more mature platform might be a better solution. Despite this issue, the competition proved to be a valuable experience for teams. Teams were able to to directly relate their knowledge of machine learning and signal processing to a real-life biomedical application. Teams also felt they were able to understand the challenges of applying machine learning to such applications, where creative approaches were taken to account for the varying granularity levels amongst biomarkers. Teams also felt the VIP Cup experience improved team members' written and oral communication skills, as well as leadership abilities. In particular, Team Synapse notes the following for the VIP Cup challenge: \say{All in all, VIP Cup 2023 was a monumental experience for our team, and thanks once again.}
%participating teams noted that the VIP Cup challenge was monumental, serving as a fun and educational experience to those involved.
\section{Conclusion}
One of the major advantages of the VIP Cup is encouraging undergraduate students to tackle real-world challenges under the support of a team of organizers and mentors within their own institution as well as between a variety of institutions. In this iteration of the VIP Cup, students were tasked with developing a machine learning algorithm to detect biomarkers of disease within an OCT clinical trial setting. Specifically, we challenged teams to find a balance between generalization and personalization for biomarker detection in ophthalmology data, encouraging teams to learn more about the complexities of medical data and how deep learning needs to adapt to such obstacles. Below, we highlight a summary of the takeaways from this competition:

\begin{itemize}
    \item Noticing which architectures perform better at local versus global feature detection and ensembling these models presents an effective approach to the generalization and personalization challenge present in medical data.
    \item Effective approaches for these tasks are also achieved through simpler methods, such as adopting a model backbone and adding sufficient hyperparameter tuning and other changes within the model architecture.
    \item Some biomarkers remain a challenge to detect overall, such as B1, where none of the top three finalists achieved the highest score for detection of this biomarker.
    \item A reliable platform for competition hosting must be chosen, including backup plans to account for the high volume of traffic.
\end{itemize}

Overall, this competition gave undergraduates an opportunity to learn about how artificial intelligence can be a powerful tool for the medical field, as well as the unique challenges one faces when applying machine learning to biomedical data. 

%due to the granularity differences that exist amongst different biomarkers as well as how differently the disease pathology presents and changes over time across patients. 
\section{Acknowledgments}
We would like to thank the judges, Lucas Thomaz, Kenneth Kin Man Lam, Mohit Prabhushankar and Ghassan AlRegib, for their time and feedback during this process. We would also like to extend thanks to Jaqueline Rash, who assisted with the competition website and helped arrange the logistics of the competition. This work was partially funded by FCT/MCTES through national funds and, when applicable, co-funded by EU funds under the projects UIDB/50008/2020 and LA/P/0109/2020. This material is also based upon work supported by the National Science Foundation Graduate Research Fellowship under Grant No. DGE-2039655. Any opinion, findings, and conclusions or recommendations expressed in this material are those
of the authors(s) and do not necessarily reflect the views of the National Science Foundation.
\section{Authors}
{\bf\em Ghassan AlRegib} (alregib@gatech.edu) obtained his Ph.D. in Electrical Engineering from the Georgia Institute of Technology (Georgia Tech). He is currently the John and Marilu McCarty Chair Professor in the School of Electrical and Computer Engineering at Georgia Tech. In the Omni Lab for Intelligent Visual Engineering and Science (OLIVES), he and his group work on robust and interpretable machine learning algorithms, uncertainty and trust, and human in the loop algorithms. He has been issued several U.S. patents and invention disclosures. He is a Fellow of the IEEE. Prof. AlRegib is active in the IEEE. He served on the editorial board of several transactions and served as the TPC Chair for ICIP 2020, ICIP 2024, and GlobalSIP 2014. He was the area editor for the IEEE Signal Processing Magazine. In 2008, he received the ECE Outstanding Junior Faculty Member Award. In 2017, he received the 2017 Denning Faculty Award for Global Engagement. He and his students received the Best Paper Award in ICIP 2019. He is a Fellow of IEEE.

{\bf\em Mohit Prabhushankar} (mohit.p@gatech.edu) received his Ph.D. degree in electrical engineering from the Georgia Institute of Technology (Georgia Tech), Atlanta, Georgia, 30332, USA, in 2021. He is currently a Postdoctoral Researcher and Teaching Fellow in the School of Electrical and Computer Engineering at the Georgia Institute of Technology in the Omni Lab for Intelligent Visual Engineering and Science (OLIVES) lab. He is working in the fields of image processing, machine learning, explainable and robust AI, active learning, and healthcare. He is the recipient of the Best Paper award at ICIP 2019 and Top Viewed Special Session Paper Award at ICIP 2020. He is the winner of the Roger P Webb ECE Graduate Research Excellence award in 2022. He is an IEEE Member.

{\bf\em Kiran Kokilepersaud} (kpk6@gatech.edu) obtained a B.S. degree in Electrical Engineering from the University of Maryland. He is currently a Ph.D. student in electrical and computer engineering at the Georgia Institute of Technology (Georgia Tech), Atlanta, Georgia, 30332, USA. He is now a Graduate Research Assistant in the School of Electrical and Computer Engineering at the Georgia Institute of Technology in the Omni Lab for Intelligent Visual Engineering and Science (OLIVES) lab. He is a recipient of the Georgia Tech President's Fellowship for excellence amongst incoming Ph.D. students. His research interests include digital signal and image processing, machine learning, and its associated applications within the medical field. He is an IEEE Student Member.

{\bf\em Prithwijit Chowdhury} (pchowdhury6@gatech.edu) received his B.Tech. degree from KIIT University, India in 2020. He joined Georgia Institute of Technology as an MS student in the department of Electrical and Computer Engineering in 2021. He is currently pursuing his Ph.D. degree as a researcher in The Center for Energy and Geo Processing (CeGP) as a member of the Omni Lab for Intelligent Visual Engineering and Science (OLIVES). His interests lie in the areas of digital signal and image processing and machine learning with applications to geophysics. He is an IEEE Student Member.

{\bf\em Zoe Fowler} (zfowler3@gatech.edu) received her B.S. degree in Electrical Engineering from Mississippi State University. She is now a PhD student in electrical and computer engineering at Georgia Institute of Technology (Georgia Tech), Atlanta, Georgia, 30332, USA. She is currently a researcher in the Omni Lab for Intelligent Visual Engineering and Sciences (OLIVES). She is a recipient of the National Science Foundation Graduate Research Fellowship Program (NSF GRFP) that recognizes graduate students based on their research and academic achievements, as well as the Georgia Tech President’s Fellowship. Her interests lie in the areas of digital signal and image processing and machine learning with applications to healthcare. She is an IEEE Student Member.

{\bf\em Stephanie Trejo Corona} is a clinical research fellow at Retina Consultants of Texas. She graduated from Rice University with a B.S. in Biochemistry and Cell Biology and a B.A. in Kinesiology in 2021. She has previously worked in the fields of synthetic and plant biology, with a passion for translational research. She was awarded the Rice University Biosciences Department Distinction in Research and Creative Work as well as multiple recognitions for her research presentations during her undergraduate career. She is currently working in the field of ophthalmology and hopes to improve healthcare access and clinical outcomes through prospective clinical trials, retrospective cohort studies, and applications of artificial intelligence.

{\bf\em Lucas Thomaz} (lucas.thomaz@co.it.pt) obtained his Ph.D. in Electrical Engineering from Universidade Federal do Rio de Janeiro. He is currently a researcher at Instituto de Telecomunicações and an associate professor in the School of Technology and Management of Polytechnic of Leiria, Leiria 2411-901, Portugal. He serves as a member of the Education Board. He is also a member of the Student Services Committee of the IEEE Signal Processing Society, supporting the Video and Image Processing Cup and the IEEE Signal Processing Cup, and the chair of the Engagement and Career Training Subcommittee. His research interests include image and video processing. He is an associate editor for IEEE Open Journal of Signal Processing. He is a Senior Member of IEEE.

{\bf\em Angshul Majumdar} (angshul@iiitd.ac.in) is Professor at TCG CREST, Kolkata. Prior to that he was a Professor at Indraprastha Institute of Information Technology, New Delhi from 2012 to 2023. He completed his PhD from the University of British Columbia in Electrical and Computer Engineering. Angshul is currently the director of the student services committee with IEEE Signal Processing Society. He has previously served the society as chair of chapter’s committee (2016-18), chair of education committee (2019) and member-at-large of the education board (2020). He is an associate editor for IEEE Open Journal of Signal Processing and Elsevier Neurocomputing. In the past, he was an associate editor for IEEE Transactions on Circuits and Systems for Video Technology. His research interests include problems in biomedical signal processing and imaging. He is a Senior Member of IEEE.
\renewcommand\refname{References}
\bibliography{main}
% I prefer to use the IEEE bibliography style. 
% That's  NOT required by the NSF guidelines. 
% Feel Free to use whatever style you prefer
\bibliographystyle{IEEEtran}
\end{spacing}
\end{document}